\documentclass[conference, letterpaper]{IEEEtran}
\IEEEoverridecommandlockouts

\usepackage{cite}
\usepackage{amsmath,amssymb,amsfonts}
\usepackage{graphicx}
\usepackage{siunitx}
\usepackage[hidelinks]{hyperref}
\usepackage{bm}
\usepackage{tikz}
\usetikzlibrary{arrows.meta,positioning,calc,fit,shapes.geometric,backgrounds,shapes.misc}
\usepackage{xcolor}
\usepackage{subcaption} 
\usepackage{algorithm}
\usepackage{algpseudocode}

\begin{document}

\title{Experimental Demonstration of Snapshot Differential Positioning with LEO Satellites}

\author{
    \IEEEauthorblockN{Soham Desai}
    \IEEEauthorblockA{\textit{Skylo Technologies} \\
    soham@skylo.tech}
    \and
    \IEEEauthorblockN{Dr. Dave Cade}
    \IEEEauthorblockA{\textit{Stanford University} \\
    davecade@stanford.edu}
}

\maketitle

\begin{abstract}
Positioning using Global Navigation Satellite Systems (GNSS) typically requires several seconds of continuous signal reception from satellites in Medium Earth Orbit (MEO). This requirement poses challenges for applications where receivers can only capture signals intermittently or operate under constrained power and visibility conditions. In such scenarios, maintaining continuous tracking or reliable line-of-sight to GNSS satellites may be difficult, and conventional GNSS frequencies may also be vulnerable to interference or jamming. 

Low Earth Orbit (LEO) satellite constellations provide an attractive alternative due to their lower orbital altitudes, which result in higher received signal strengths, as well as their operation across a wide range of spectrum including Mobile-Satellite Service (MSS) and terrestrial L and S bands. These characteristics make LEO signals promising for navigation in challenging environments.

This work presents a snapshot-based differential positioning framework that leverages signals from LEO satellites. In the proposed approach, a receiver collects signals for short durations (5–10 seconds) before entering a low-power state, enabling positioning with intermittent observations. Doppler measurements from multiple satellites are combined with a differential measurement model using a fixed reference receiver to mitigate common errors such as satellite clock bias and ephemeris uncertainty. Experimental results demonstrate that the proposed differential Doppler framework operates effectively within the constraints of snapshot-based reception. The method achieves a position error reduction of approximately 47\% even when only three satellites are simultaneously visible to both the rover and the reference station.
\end{abstract}


\section{Introduction}

Reliable and rapid positioning and tracking are critical across many scientific and industrial applications \cite{Yang2024}. However, in some real-world scenarios—such as tracking marine mammals that surface briefly—Global Navigation Satellite Systems (GNSS) struggle to provide consistent solutions. Baleen whales, for example, may surface for only 3–5 seconds, which is often insufficient for conventional GNSS receivers to acquire a position fix. Although technologies such as Fastloc-GPS have improved acquisition times for wildlife tracking \cite{Dujon2014}, they still rely on signals from Medium Earth Orbit (MEO) satellites, which arrive at the Earth’s surface with relatively low power and are susceptible to attenuation in challenging environments.

Low Earth Orbit (LEO) satellite constellations have recently emerged as a promising alternative. Operating at altitudes between 400 km and 1500 km, LEO satellites can provide received signal strengths roughly 30 dB higher than GNSS \cite{Neinavaie2022}. Their rapid orbital motion also produces significant Doppler dynamics and changing geometry that can be exploited for positioning using Signals of Opportunity (SoOP). Recent studies have demonstrated the feasibility of navigation using commercial constellations such as Starlink, OneWeb, and Iridium \cite{Kassas2023}, with Doppler-based frameworks achieving accuracies on the order of tens of meters \cite{Orabi2021}.

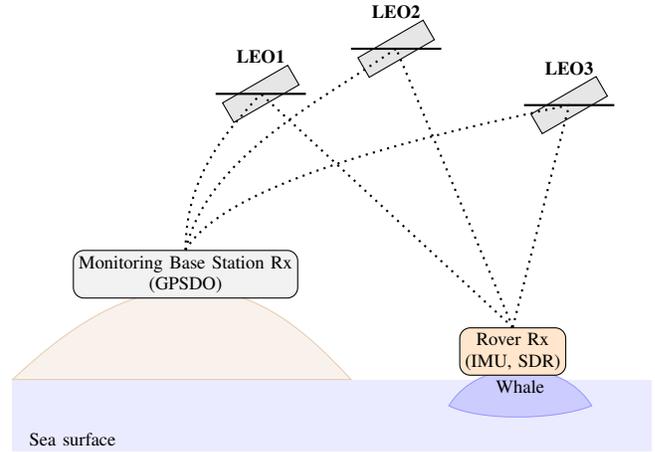
\begin{figure}[t]
\centering
\begin{tikzpicture}[font=\scriptsize,>=Latex,scale=1, every node/.style={transform shape}]
\tikzset{
  satellite/.style = {draw,rotate=30,minimum width=10mm,minimum height=3mm,fill=gray!20},
  tagbox/.style    = {draw,rounded corners,fill=orange!20,align=center,inner sep=2pt},
  refbox/.style    = {draw,rounded corners,fill=gray!10,align=center,inner sep=2pt},
  linkSoOP/.style  = {dotted,thick},
}
\fill[blue!8] (-4.5,-1.55) rectangle (4,-2.5);
\node[anchor=west] at (-4.4,-2.35) {Sea surface};
\draw[fill=brown!10,draw=brown!40] (-4.5,-1.55)
  .. controls (-3.8,-0.9) and (-3.0,-0.3) ..  (-2.0,-0.35)
  .. controls (-1.2,-0.4) and (-0.6,-0.9) ..  (0.0,-1.55) -- cycle;
\node[refbox] (base) at (-2.2,-0.15) {Monitoring Base Station Rx\\(GPSDO)};
\draw[fill=blue!20,draw=blue!40] (1.3,-1.9) .. controls (1.7,-1.25) and (2.8,-1.25) .. (3.2,-1.9)
 .. controls (2.7,-2.1) and (1.7,-2.1) .. cycle;
\node at (2.25,-1.65) {Whale};
\node[tagbox,anchor=south] (tag) at (2.15,-1.5) {Rover Rx\\(IMU, SDR)};
\foreach [count=\i] \x/\y in {-1.2/2.25, 0.6/2.85, 2.9/2.1}{
  \node[satellite] (sat\i) at (\x,\y) {};
  \draw[thick] (\x-0.6,\y) -- (\x+0.6,\y);
\node[above=8pt,font=\scriptsize\bfseries] at (\x,\y) {LEO\i};
}
\draw[linkSoOP] (tag.north) -- (0.6,2.85);
\draw[linkSoOP] (tag.north) -- (-1.2,2.25);
\draw[linkSoOP] (tag.north) -- (2.9,2.1);
\draw[linkSoOP] (base.north) .. controls (-2.2,1.2) and (-1.6,1.8) .. (-1.2,2.25);
\draw[linkSoOP] (base.north) .. controls (-1.9,1.4) and (-0.5,2.1) .. (0.6,2.85);
\draw[linkSoOP] (base.north) .. controls (-1.7,1.0) and (0.5,1.6) .. (2.9,2.1);
\end{tikzpicture}
\caption{Positioning of briefly surfacing whales using snapshot measurement of LEO satellites}
\label{fig:concept}
\end{figure}

Despite these advantages, using commercial LEO satellites for positioning presents several challenges. Unlike GNSS satellites, many LEO spacecraft lack highly stable atomic clocks, and precise ephemerides are often unavailable. Navigation algorithms must therefore rely on Two-Line Element (TLE) sets and simplified orbital propagation models, which can introduce satellite position and velocity errors on the order of several kilometers. Differential Doppler positioning addresses this limitation by using a reference station to estimate and cancel common-mode errors such as atmospheric effects and ephemeris inaccuracies, significantly improving accuracy \cite{Hayek,Baseline2023}.

Most prior work assumes receivers capable of continuous tracking. In contrast, many sensing platforms operate under strict power constraints or intermittent signal visibility. This paper investigates \textit{snapshot} positioning, where the receiver collects signals only during short observation windows. We present experimental results for a snapshot differential Doppler framework using the Iridium constellation. As illustrated in Fig.~\ref{fig:concept}, a fixed monitoring base station provides differential corrections to estimate the unknown position of a mobile rover tag.

\begin{figure*}[htbp]
    \centering
    
    \begin{minipage}[c]{0.64\textwidth}
        \centering
        
        \resizebox{\linewidth}{!}{
           \begin{tikzpicture}[
                >=Stealth,
                node distance=0.8cm and 2.5cm,
                box/.style={draw, rectangle, rounded corners, minimum width=3.8cm, minimum height=0.8cm, align=center, fill=blue!5, font=\small},
                sat/.style={draw, ellipse, minimum width=2.5cm, minimum height=0.8cm, align=center, fill=orange!10, font=\footnotesize},
                data/.style={draw, chamfered rectangle, minimum width=3.8cm, minimum height=0.8cm, align=center, fill=green!5, font=\small},
                arrow/.style={->, thick}
            ]
            
            \node[sat] (sat2) {Common Satellite 1\\(e.g., Satellite 67)};
            \node[sat, left=0.5cm of sat2] (sat1) {Common Satellite 2\\(e.g., Satellite 25)};
            \node[sat, right=0.5cm of sat2] (sat3) {Common Satellite 3};
            \node[draw, dashed, inner sep=0.3cm, fit=(sat1) (sat2) (sat3), label=above:\textbf{Space Segment (Identical TLE Errors)}] (space) {};
            
            \node[box, below left=1.5cm and 0.2cm of sat2, fill=red!10] (base) {\textbf{Monitoring Base Station}\\Known Location};
            \node[data, below=of base] (base_meas) {Measured Doppler\\$f_{\text{meas}, B}(t)$};
            \node[box, below=of base_meas] (calc_theo) {Calculate Theoretical\\$f_{\text{theo}, B}(t)$};
            \node[box, below=of calc_theo] (error_calc) {Compute Residuals\\$e(t)$};
            
            \node[box, below right=1.5cm and 0.2cm of sat2, fill=red!10] (rover) {\textbf{Rover Station}\\Unknown Location};
            \node[data, below=of rover] (rover_meas) {Measured Doppler\\$f_{\text{meas}, U}(t)$};
            
            \node[box] at (rover_meas |- error_calc) (apply_corr) {Apply Differential Corrections\\$f_{\text{corr}, U} = f_{\text{meas}, U} - e(t)$};
            
            \node[box, below=0.8cm of apply_corr] (init_est) {\textbf{Initial Position Estimation}\\Zero Doppler Shift ($t_{f_D=0}$) $\rightarrow \vec{r}_U$};
            
            \node[box, below=1cm of init_est, fill=white] (trial_state) {\textbf{Trial State Vector ($\vec{s}$)}\\$\vec{s} = (\phi_U, \lambda_U, h_U, f_O, \dot{f}_O)$};
            \node[box, below=0.6cm of trial_state, fill=white] (trial_curve) {\textbf{Trial Doppler Curve ($C_T$)}\\$f_D = \frac{1}{c} f_B \dot{\rho} + f_O + \dot{f}_O t$};
            \node[box, below=0.6cm of trial_curve, fill=white] (optimizer) {Iterative Curve Fitting\\Compare $C_T$ with $f_{\text{corr}, U}$};
            
            \node[draw, thick, dotted, inner sep=0.3cm, fill=gray!5, fit=(trial_state) (trial_curve) (optimizer), label={[font=\bfseries]above:Position Solver                  (            - Iterative Routine)}] (solver_box) {};
            
            \node[box, fill=white] at (trial_state) {\textbf{Trial State Vector ($\vec{s}$)}\\$\vec{s} = (\phi_U, \lambda_U, h_U, f_O, \dot{f}_O)$};
            \node[box, fill=white] at (trial_curve) {\textbf{Trial Doppler Curve ($C_T$)}\\$f_D = \frac{1}{c} f_B \dot{\rho} + f_O + \dot{f}_O t$};
            \node[box, fill=white] at (optimizer) {Iterative Curve Fitting\\Compare $C_T$ with $f_{\text{corr}, U}$};
            
            \node[data, below=0.8cm of solver_box, fill=blue!10] (final_pos) {\textbf{Final Position Estimate}\\Optimized $\vec{s}$};
            
            \draw[arrow, dashed] (sat1) -- (base);
            \draw[arrow, dashed] (sat2) -- (base);
            \draw[arrow, dashed] (sat3) -- (base);
            \draw[arrow, dashed] (sat1) -- (rover);
            \draw[arrow, dashed] (sat2) -- (rover);
            \draw[arrow, dashed] (sat3) -- (rover);
            
            \draw[arrow] (base) -- (base_meas);
            \draw[arrow] (base_meas) -- (calc_theo);
            \draw[arrow] (calc_theo) -- (error_calc);
            
            \draw[arrow] (rover) -- (rover_meas);
            \draw[arrow] (rover_meas) -- (apply_corr);
            \draw[arrow] (apply_corr) -- (init_est);
            \draw[arrow] (init_est) -- (trial_state);
            
            \draw[arrow] (trial_state) -- (trial_curve);
            \draw[arrow] (trial_curve) -- (optimizer);
            
            \draw[arrow, dashed, color=blue] (optimizer.west) -- ++(-0.6,0) |- node[pos=0.25, left, font=\scriptsize, align=center] {Update\\State Vector} (trial_state.west);
            
            \draw[arrow] (optimizer) -- (final_pos);
            
            \draw[arrow, thick, color=red] (error_calc.east) -- node[above, align=center, font=\footnotesize, text=red] {Interpolated\\Error $e(t)$} (apply_corr.west);
            \end{tikzpicture}
        }
        \caption{Complete end-to-end architecture of the Differential Doppler Positioning framework}
        \label{fig:diff_doppler_arch_complete}
    \end{minipage}%
    \hfill
    \begin{minipage}[c]{0.32\textwidth}
        \centering
        
        \begin{subfigure}[b]{\linewidth} 
            \centering
            \includegraphics[width=\linewidth]{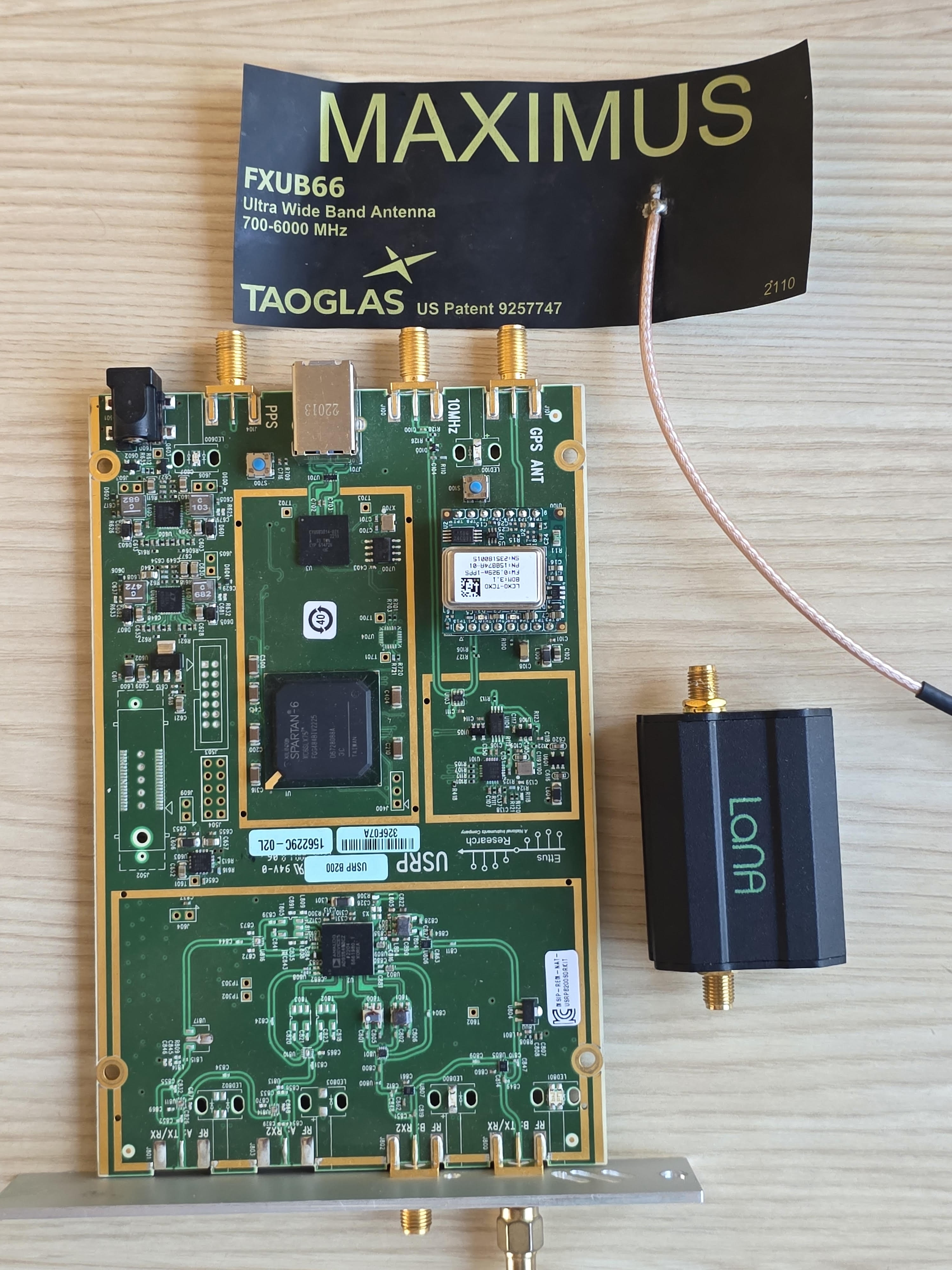}
            \caption{Monitoring Base Station}
            \label{fig:fixed_setup}
        \end{subfigure}
        
        \vspace{0.8cm} 
       
        \begin{subfigure}[b]{\linewidth}
            \centering
            \includegraphics[angle=270, origin=c, width=\linewidth]{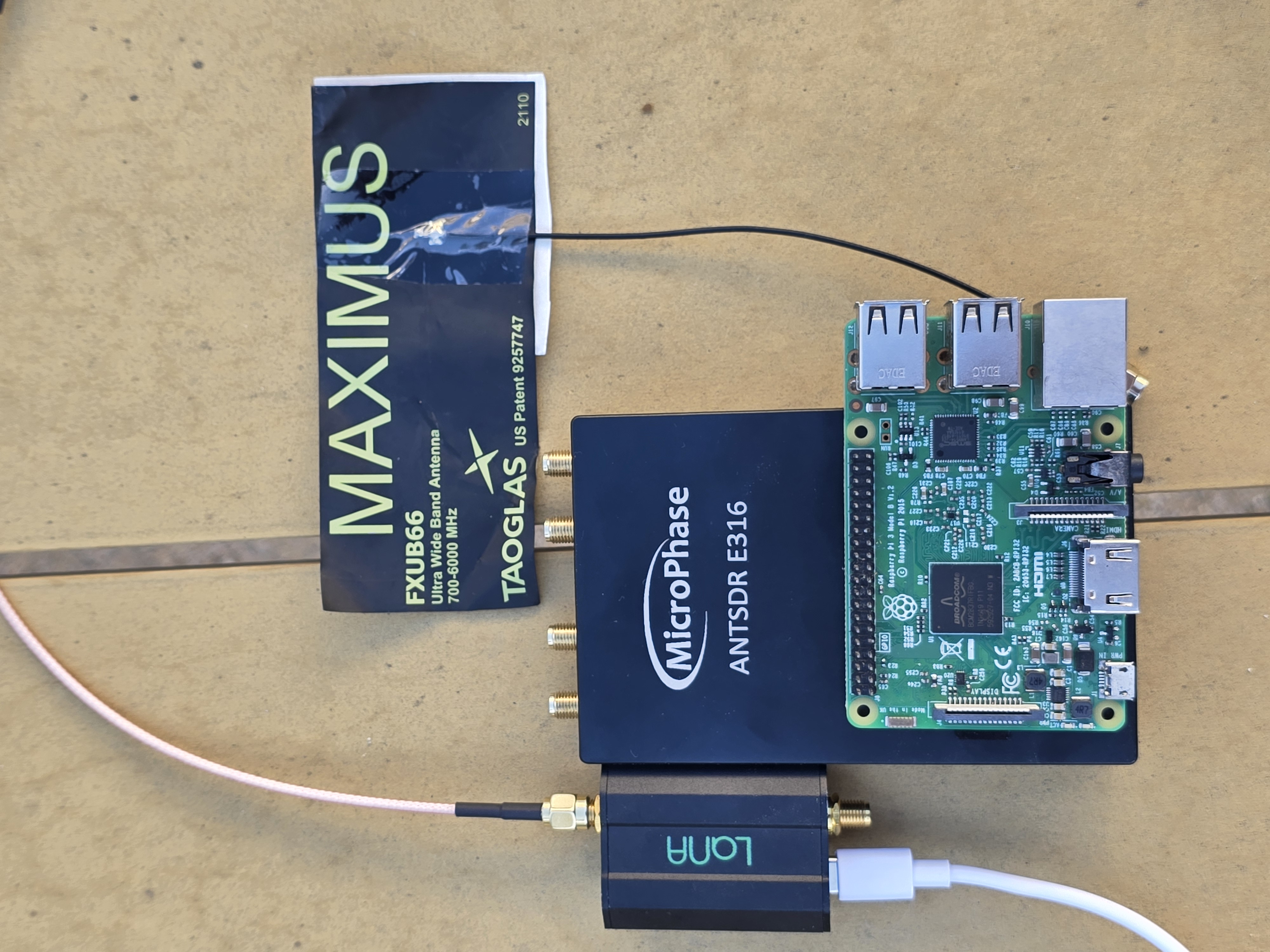}
            \caption{Rover Station}
            \label{fig:rover_setup}
        \end{subfigure}
        
        \caption{Hardware setup utilized during the experimental data collection.}
        \label{fig:hardware_setup}
    \end{minipage}
    
\end{figure*}

\section{Positioning Methodology}

The position estimation method detailed in this work is the Doppler-curve-fitting method. The process involves Satellite Matching, Initial Position Estimation, Actual Position Finding, and applying Differential Corrections from measurements done at a Monitoring Base station.

\subsection{Pre-Processing \& Initial Estimation}
A critical challenge in opportunistic navigation is determining which public satellite (identified by its NORAD ID) transmitted a captured signal frame. If the mapping between the satellite's internal ID or transmission frequency and its public NORAD ID is not known, we use spatial matching methods to identify the correct match by minimizing the Euclidean distance between a satellite's self-reported position within the data frame and the predicted positions derived from public ephemerides (TLEs).

Before the main iterative search begins, the system requires a rough initial estimate of the user position ($\vec{r}_{U}$). We utilize the zero-Doppler-shift principle. The system measures and tracks the Doppler shifts from multiple passing satellites over time, generating distinct Doppler curves. The moment of zero-Doppler-shift ($t_{f_D=0}$) is interpolated. The initial position is estimated by averaging the satellite positions at these interpolated moments, as the user is located on a line perpendicular to the satellite's ground track at $t_{f_D=0}$.

\subsection{Position Solver (Doppler Curve Fitting)}
The final position is calculated using an iterative optimization routine based on the theoretical physical model mentioned below.

\subsubsection{Trial State Vector}
The navigation system estimates a state vector ($\vec{s}$) which includes the user's geodetic position and parameters to compensate for the local receiver's clock errors:
\begin{equation}
\vec{s} = (\phi_U, \lambda_U, h_U, f_O, \dot{f}_O)
\end{equation}
where $\phi_U, \lambda_U, h_U$ represent the user's latitude, longitude, and altitude; $f_O$ is the constant frequency offset (Hz) of the receiver clock; and $\dot{f}_O$ is the linear frequency drift rate (Hz/s).

\subsubsection{Trial Doppler Curve Generation}
For any trial state vector $\vec{s}$, a trial Doppler curve ($C_T$) is calculated using the underlying physical model:
\begin{equation}
f_D = \frac{1}{c} f_B \dot{\rho} + f_O + \dot{f}_O t
\end{equation}
where $\dot{\rho}$ is the range rate (relative velocity) between the satellite and the trial user position, $f_B$ is the base frequency, and $t$ is time.

\subsection{Differential Doppler Method}
To eliminate the large common-mode errors inherent to pure communication purpose LEO signals (specifically TLE-induced ephemeris inaccuracies and atmospheric delays), a differential model is applied. The fixed monitoring base station, located at known coordinates, simultaneously captures the Doppler shift from the same satellites. 

The true geometric Doppler shift and the rover's localized clock errors ($f_{\text{corr}, U}$) are obtained by subtracting the base station's isolated error from the rover's measurements:
\begin{equation}
f_{\text{corr}, U}(t) = f_{\text{meas}, U}(t) - (f_{\text{meas}, B}(t) - f_{\text{theo}, B}(t) - f_{\text{clock}, B}(t))
\end{equation}
This interpolated and corrected rover signal is then fed back into the position solver. The overall positioning system and differential architecture is detailed in Fig. \ref{fig:diff_doppler_arch_complete}.

\section{Experimental Evaluation and Results}

\subsection{Experimental Setup}
To validate the proposed snapshot differential positioning system, a field experiment was conducted in the San Francisco Bay Area using two Software-Defined Radio (SDR) stations.

\subsubsection{Monitoring Base Station (Reference)}
The reference station (Fig. \ref{fig:fixed_setup}) was equipped with an Ettus USRP B200 SDR and a GPS-Disciplined Oscillator (GPSDO) to maintain a highly stable clock reference (10 MHz PPS), minimizing local clock drift. It was connected to a laptop running Ubuntu. A Taoglas FXUB66 ultra-wideband antenna and a Nooelec Lana low noise amplifier (LNA) were used to capture the L-band Iridium signals.

\subsubsection{Rover Station}
The rover station (Fig. \ref{fig:rover_setup}) was located at a separate location, establishing a baseline distance of approximately 10 km from the base station. Designed to mimic a low-cost, power-constrained marine tag, it utilized a MicroPhase ANTSDR E316. This was connected to a Raspberry Pi and similar antenna and LNA connection were used. Unlike the fixed station, the rover did not have a GPSDO, instead relying on its internal oscillator.

\subsection{Data Collection and Results}

\subsubsection{Snapshot Data Collection Profile}
To simulate a reciever only active for a few seconds, the Rover did not track satellites continuously. Instead, it collected data in distinct bursts over a duration of approximately 55 minutes, with an average burst duration of 5 seconds. During this collection window, the datasets shared three overlapping satellites which provided data needed for differential corrections.

\begin{figure*}[htbp] 
    \centering 
    \includegraphics[angle=0, width=\linewidth]{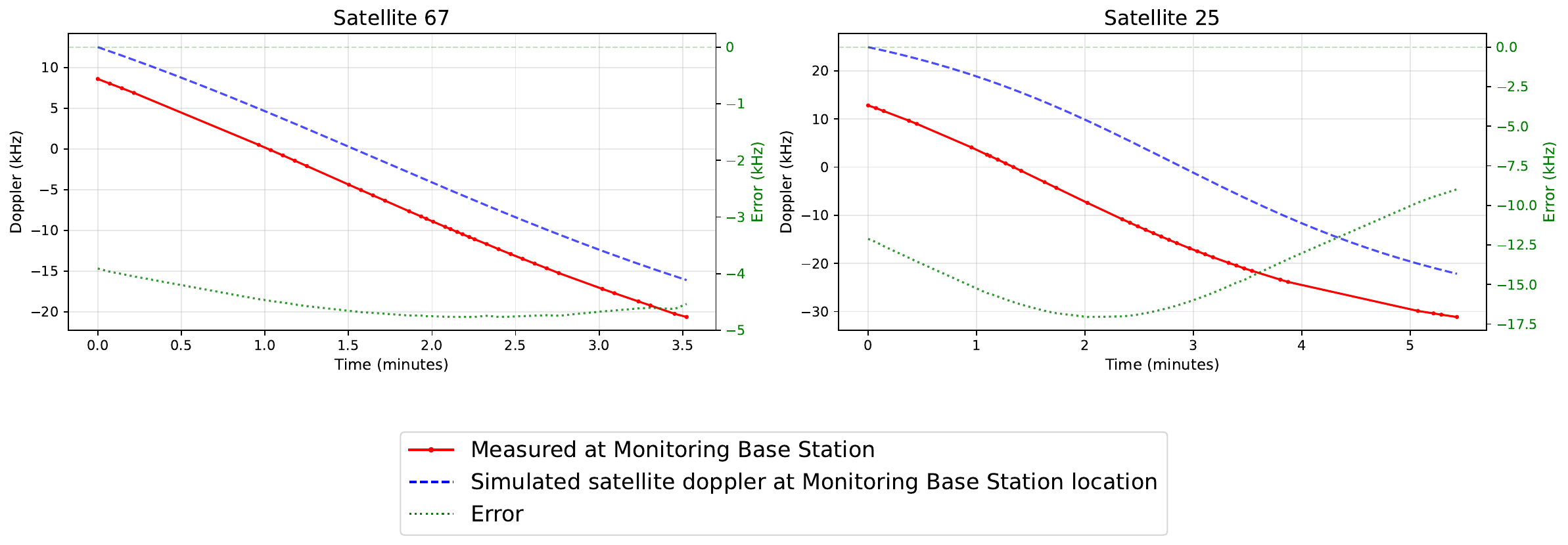} 
    \caption{Doppler plots showing Measured, Simulated and Error Results}
    \label{fig:matching} 
\end{figure*}

\subsubsection{Signal Quality Analysis}
A comparison of Doppler measurement quality highlights the challenge of using uncompensated hardware in a snapshot scenario. The monitoring base station, aided by the GPSDO, continuous tracking, and an unobstructed view, achieved a highly stable average root mean square error (RMSE) of just 5-7 Hz when comparing with the simulated doppler using satellites TLE files. In contrast, the rover exhibited an average RMSE of 80-100 Hz. This increased noise is attributed to the start-stop nature of the snapshot acquisition and the thermal drift of the uncompensated crystal oscillator. These inherent hardware constraints make standalone positioning highly inaccurate, necessitating the differential Doppler mechanism.

\subsection{Positioning Performance}
The performance of the snapshot differential positioning system was evaluated by comparing the estimated positions against the ground truth.

\begin{figure}[htbp]
\centering
\includegraphics[width=\columnwidth, keepaspectratio]{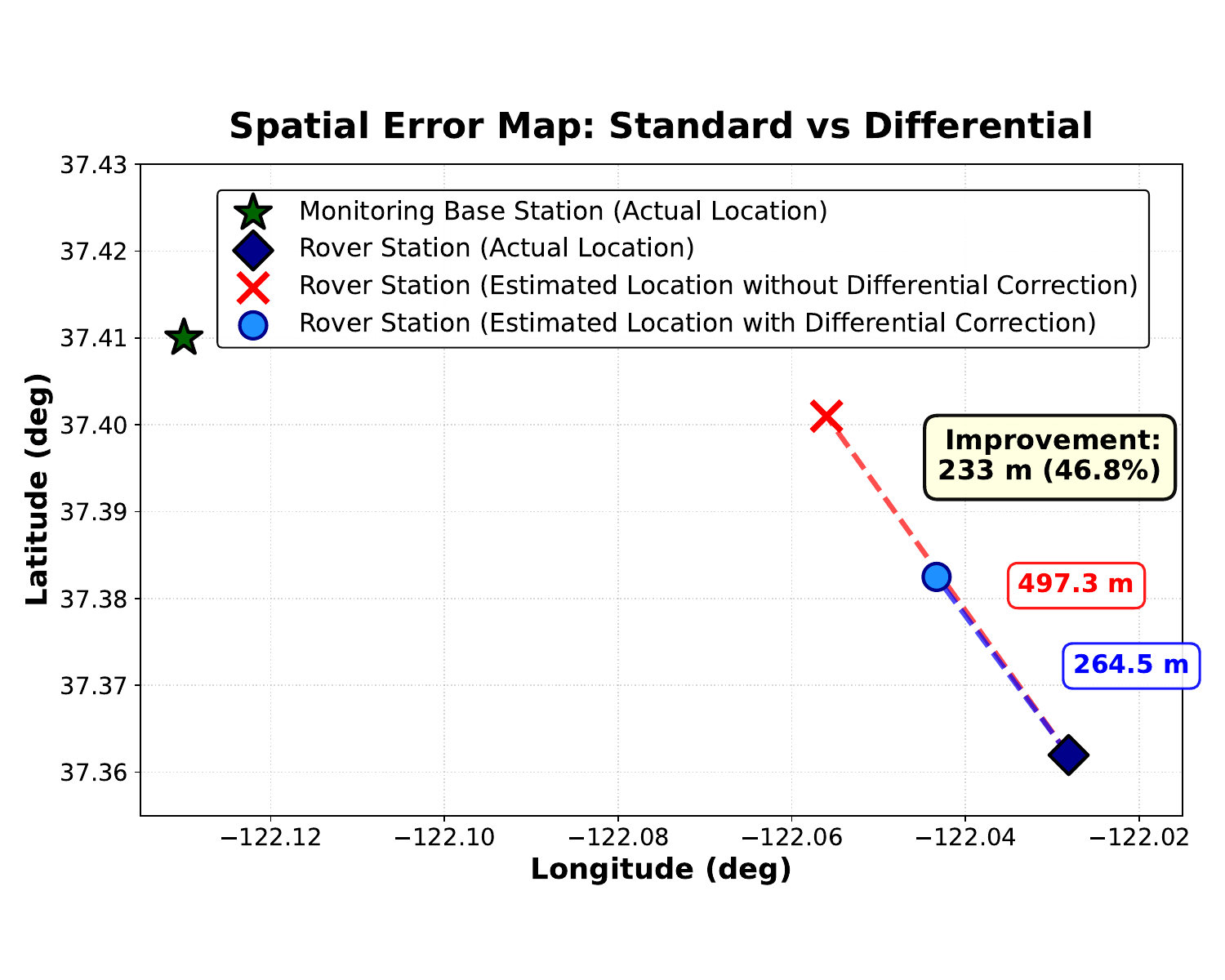}
\caption{Experiment results}\label{fig:results_map}
\end{figure}

\begin{itemize}
\item \textbf{Rover Standalone (Standard):} When the rover position was estimated using standard Doppler processing (without differential corrections), the result deviated from the true position by approximately 500 m. This large initial error is overwhelmingly driven by imprecise TLE orbits coupled with the receiver's significant clock drift.
\item \textbf{Rover Differential:} By applying the time-interpolated differential corrections derived from the fixed station, the estimated position improved drastically. The absolute error was reduced to less than 265 m, with the final RMS error dropping to 10-20 Hz (see Fig. \ref{fig:results_map}).
\end{itemize}

\section{Conclusion}
This work experimentally demonstrated a snapshot differential positioning framework using LEO satellites, specifically addressing the constraints of applications where continuous tracking is impossible, such as tracking assets that only emerge for brief moments. By leveraging a reference station to calculate time-interpolated differential Doppler corrections, we successfully mitigated the large errors caused by TLE inaccuracies and atmospheric delays. Initial experimental results showed a dramatic positioning error reduction of approximately 47\%, pulling standalone errors of approximately 500 m down to under 265 m. This was achieved using brief 5-second data snapshots utilizing just three overlapping satellites. This confirms that differential LEO navigation is a viable, highly impactful solution for position estimation and tracking in environments where continuous sky visibility is severely limited. Positioning results will be improved with more than 3 overlapping satellites and the integration of other sensors.

While highly effective, one shortcoming of the current implementation is the requirement for spatial proximity; the Monitoring Station and Rover must be relatively close. In future work, we plan to explore the integration of Terrestrial Networks (TN) and Non-Terrestrial Networks (NTN). By utilizing widespread TN cellular base stations as distributed reference nodes to continuously monitor satellite signals, we can overcome the strict spatial limitations of a single base station, enabling large-scale, ubiquitous passive radar and positioning applications. Finally, the use of Inertial Measurement Unit (IMU) sensor fusion will be explored to bridge positioning gaps during the periods between brief signal acquisitions.

\section{Acknowledgment}
The authors thank Dr. Andrew Nuttal, Meghna Agarwal, Dr. Antonio Albanese, and anonymous reviewers for their careful reading and helpful comments, which have significantly improved the quality of this manuscript.

\bibliographystyle{IEEEtran}

\begin{thebibliography}{99}

\bibitem{Yang2024}
Y. Yang \emph{et al.}, ``Positioning Using Wireless Networks: Applications, Recent Progress and Future Challenges,'' 2024.

\bibitem{Dujon2014}
A. M. Dujon \emph{et al.}, ``The accuracy of Fastloc-GPS locations and implications for animal tracking,'' \emph{Methods in Ecology and Evolution}, 2014.

\bibitem{Neinavaie2022}
M. Neinavaie \emph{et al.}, ``First results of differential Doppler positioning with unknown Starlink satellite signals,'' in \emph{Proc. IEEE Aerospace Conf.}, 2022.

\bibitem{Kassas2023}
Z. M. Kassas \emph{et al.}, ``Navigation with Multi-Constellation LEO Satellite Signals of Opportunity: Starlink, OneWeb, Orbcomm, and Iridium,'' in \emph{IEEE/ION PLANS}, 2023.

\bibitem{Orabi2021}
M. Orabi \emph{et al.}, ``Opportunistic Navigation with Doppler Measurements from Iridium Next and Orbcomm LEO Satellites,'' \emph{IEEE}, 2021.

\bibitem{sgp4}
F. R. Hoots and R. L. Roehrich, ``Spacetrack report no. 3: Models for propagation of NORAD element sets,'' \emph{U.S. Air Force Aerospace Defense Command}, 1980.

\bibitem{Hayek}
S. Hayek \emph{et al.}, ``Assessment of Differential Doppler Navigation with Starlink LEO Satellite Signals of Opportunity,'' \emph{Tech. Report}, 2023.

\bibitem{Baseline2023}
C. Zhao \emph{et al.}, ``Analysis of Baseline Impact on Differential Doppler Positioning and Performance Improvement Method for LEO Opportunistic Navigation,'' \emph{IEEE Transactions on Instrumentation and Measurement}, 2023.

\bibitem{iridiumref}
M. Joerger \emph{et al.}, ``Iridium signal classification and decoding for navigation,'' \emph{Proceedings of the ION GNSS+}, 2018.

\bibitem{voskaDP}
V. Voska, ``DP: Differential Positioning Repository.'' [Online]. Available: https://github.com/voskavoj/DP

\end{thebibliography}

\end{document}